\documentclass[twocolumn]{aastex63}
\usepackage{lineno}
\usepackage[utf8]{inputenc}
\usepackage{amsmath}
\usepackage{amssymb}
\usepackage{graphicx}
\usepackage{epstopdf}
\usepackage{hyperref}
\usepackage{times}
\usepackage{verbatim, longtable}	
\usepackage{rotating,multirow} 
\usepackage{CJKutf8}

\newcommand{\D}{\displaystyle}

\def\oiii{[{O~\sc III}]}
\def\feii{{Fe~\sc II}}

\shorttitle{jetted AGNs}
\shortauthors{Chen et al.}

\begin{document}
\title{Curvature of the spectral energy distribution, Compton dominance and synchrotron peak frequency in jetted AGNs}

\correspondingauthor{Yongyun Chen}
\email{ynkmcyy@yeah,net}

\correspondingauthor{Qiusheng Gu}
\email{qsgu@nju.edu.cn}

\author{Yongyun Chen$^{*}$ \begin{CJK*}{UTF8}{gkai}(陈永云)\end{CJK*}}
\affiliation{College of Physics and Electronic Engineering, Qujing Normal University, Qujing 655011, P.R. China}

\affiliation{Key Laboratory of Modern Astronomy and Astrophysics (Nanjing University), Ministry of Education, Nanjing 210093, China}

\author{Qiusheng Gu$^{*}$ \begin{CJK*}{UTF8}{gkai}(顾秋生)\end{CJK*}}
\affiliation{School of Astronomy and Space Science, Nanjing University, Nanjing 210093, P. R. China}

\affiliation{Key Laboratory of Modern Astronomy and Astrophysics (Nanjing University), Ministry of Education, Nanjing 210093, China}

\author{Junhui Fan \begin{CJK*}{UTF8}{gkai}(樊军辉)\end{CJK*}}
\affiliation{Center for Astrophysics,Guang zhou University,Guang zhou510006, China}

\author{Xiaoling Yu\begin{CJK*}{UTF8}{gkai}(俞效龄)\end{CJK*}}
\affiliation{College of Physics and Electronic Engineering, Qujing Normal University, Qujing 655011, P.R. China}

\author{Nan Ding \begin{CJK*}{UTF8}{gkai}(丁楠)\end{CJK*}}
\affiliation{School of Physical Science and Technology, Kunming University 650214, P. R. China}

\author{Dingrong Xiong \begin{CJK*}{UTF8}{gkai}(熊定荣)\end{CJK*}}
\affiliation{Yunnan Observatories, Chinese Academy of Sciences, Kunming 650011,China}

\author{Xiaotong Guo \begin{CJK*}{UTF8}{gkai}(郭晓通)\end{CJK*}}
\affiliation{Anqing Normal University, 246133, P. R. China}

\begin{abstract}
We collect a large sample with a reliable redshift detected by the Fermi satellite after 10 years of data (4FGL-DR2), including blazars, $\gamma$-ray Narrow-line Seyfert 1 galaxies ($\gamma$NLS1s), and radio galaxies. The spectral energy distributions (SEDs) of these Fermi sources are fitted by using a second-degree polynomial, and some important parameters including spectral curvature, synchrotron peak frequency, and peak luminosity are obtained. Based on those parameters, we discuss the Fermi blazar sequence and the particle acceleration mechanism. Our main results are as follows:(i) By studying the relationship between the synchrotron peak frequency and the synchrotron peak frequency luminosity, jet kinetic power, and $\gamma$-ray luminosity for jetted AGNs, we find an ``L'' shape in the Fermi blazar sequence. (ii) There is a significant anti-correlation between Compton dominance, black hole spin, and the synchrotron peak frequency for jetted AGNs, respectively. These results support that the $\gamma$NLS1s and radio galaxies belong to the Fermi blazar sequence. (iii) On the basis of previous work, statistical or stochastic acceleration mechanisms can be used to explain the relationship between synchrotron peak frequency and synchrotron curvature. For different subclasses, the correlation slopes are different, which implies that the Fermi sources of different subclasses have different acceleration mechanisms. (iv) The FSRQs and $\gamma$NLS1s have a higher median spin of a black hole than BL Lacs and radio galaxies.  
\end{abstract}


\section{Introduction}
Blazar is a special subclass of active galactic nuclei (AGNs), whose jets point to observers. According to the equivalent width (EW) of the emission line, blazar is divided into two categories: flat spectrum radio quasars (FSRQs) and BLLac objects (BLLac). FSRQ has $EW>5$\AA, while the source of $EW<5$\AA~is BLLacs \citep{Urry1995}. Some authors proposed a physical classification, \cite{Ghisellini2011} found that the ratio of broad-line region luminosity ($L_{\rm BLR}$) to Eddington luminosity ($L_{\rm Edd}$) of FSRQs is $L_{\rm BLR}/L_{\rm Edd}\geq10^{-3}$, while BL Lacs are less than this value. The result may reflect the difference in their accretion mechanism. \cite{Foschini2017} proposed another possible classification scheme by studying the relation between jet power and the luminosity of accretion disk for blazars and $\gamma$NLS1s. They found two branches, namely, electron cooling in different environments and black hole mass, based on the main factors driving the change of jet power. According to the mass of black holes, FSRQs and BL Lacs are divided into one category, while the $\gamma$NLS1s is another. In terms of electron cooling, FSRQs and $\gamma$NLS1s are divided into one category, while BL Lacs are another.   

Seyfert galaxies are low luminosity AGNs, which are used to probe the jet, corona and disk connection \citep[e.g.,][]{Wilkins2015}. Seyfert galaxies are classified into Seyfert 1 and Seyfert 2 according to whether there are broad permitted emission lines in their optical spectra, respectively. Some Seyfert 1 galaxies show broad permitted lines with narrow widths, called
narrow-line Seyfert 1 galaxies (NLS1s). The NLS1s have a full width at half-maximum (FWHM) of $H\beta<2000~\rm km~s^{-1}$, weak \oiii~emssion line, and the presence of strong \feii~multiplets \citep[e.g.,][]{Osterbrock1985}. NLS1 is usually found to reside in spiral galaxies with low black hole mass \citep[e.g.,][]{Grupe2004}. \cite{Foschini2011} found that the jet power of FSRQs and $\gamma$NLS1s depends on the black hole mass, which suggests that the accretion disk of FSRQs and $\gamma$NLS1s was dominated by radiation pressure. Some authors found that the SED of $\gamma$NLS1s is Compton-dominated, similar to that of FSRQs \citep[e.g., ][]{Abdo2009a, Abdo2009b, Yang2015}. \cite{Foschini2015} and \cite{Berton2016} already found that flat-spectrum radio-loud narrow-line Seyfert 1 galaxies (F-RLNLS1s) might be the low-luminosity and low-mass tail of FSRQs. According to the SED modeling, \cite{Paliya2019} found that the Physical Properties of $\gamma$NLS1s are similar to FSRQs.

Radio galaxies are considered to be the parent population of blazars: while blazars are aligned to our line of sight, radio galaxies have their jets oriented at larger viewing angles. According to radio morphology, radio galaxies are historically divided into FR I and FR II radio galaxies \citep{Fanaroff1974}. FR I radio galaxies have bright jets close to the nucleus, while FR II radio galaxies show prominent hotspots far from it. Some authors claim that the morphological separation of FR I and FR II radio galaxies may be related to the formation of their jets, that is, through the Blandford-Znajek (BZ: \cite{Blandford1977}) process of extracting the spin energy of black holes, or the Blandford-Payne (BP, \cite{Blandford1982}) process of extracting the rotational energy through accretion disks. The observation of EHT (Event Horizon Telescope, \cite{EHT2019}) showed that M87 hosts an FR I jet, which appears consistent with the BZ process. According to the unified model of AGNs, FSRQs and FRII radio galaxies are unified, and BL Lacs and FR I radio galaxies are unified \citep{Urry1995}. \cite{Chen2015} found that the jet power of FSRQs and FR II radio galaxies depends on the black hole mass, while the jet power of BL Lacs and FR I radio galaxies depends on the accretion rates. These results suggest that the accretion disk of FSRQs and FR II radio galaxies is dominated by radiation pressure, and the accretion disk of BL Lacs and FR I radio galaxies is dominated by gas pressure.                         

Through the study of 126 blazars, it was found that there was a negative correlation between the luminosity of the synchrotron peak ($L_{\rm pk}^{\rm sy}$) and the synchrotron peak frequency ($\nu_{\rm pk}^{\rm sy}$) \citep{Fossati1998}. \cite{Fossati1998} also noticed anti-correlations between the 5 GHz luminosity, synchrotron peak frequency ($\nu_{\rm pk}^{\rm sy}$), and the Compton dominance (the ratio of Compton peak to synchrotron peak luminosity). These correlations are claimed
as evidence for a ``blazar sequence''. Subsequently, \cite{Ghisellini1998} proposed a physical explanation for the blazar sequence. They fitted the broad-band spectra of blazars and suggested that radiation cooling led to the emergence of the blazar sequence. The main idea is that if the radiative cooling of emitted electrons is severe, then the emitted electrons can not achieve high energies. According to this idea, in FSRQs, the broad line region and a molecular torus are essential sources of seed photons for the external Compton process, which means that the random Lorentz factor of the electrons emitting at the peaks of the SED, $\gamma_{\rm peak}$, is relatively small. Therefore, their synchrotron radiation peaks appear in the submillimeter band, while the high-energy peaks appear in the MeV band. Another important result is that the inverse Compton process becomes dominant over the synchrotron process, which means a large Compton dominance. In BL Lacs, the lack of an important thermal source implies the paucity of seed photons for Compton scattering. Radiation cooling is less severe, allowing electrons to achieve large $\gamma_{\rm peak}$. Therefore, the peak of the synchrotron is in the UV-X-ray band, and the high-energy hump peaks are in the GeV, sometimes even in the TeV band. The lack of ambient seed photons implies that only the SSC process can contribute to high energy emissions. Thus, high-energy components usually do not play a dominant role in BL Lacs. This idea can be verified by applying a leptonic one-zone model, which allows inferring the physical parameters of different types of sources. We also know that the presence or absence of broad-emitting lines is related to the luminosity of the accretion disk and its accretion regime. In addition, there is a relationship between the jet power and the luminosity of accretion disk. Therefore, the blazar sequence can also be considered as the sequence whose main parameter is the jet power or equivalent to the disk luminosity. On the other hand, we must also consider the mass of the central black hole, because when the disk luminosity is about $10^{-3}-10^{-2}$ of the Eddington one, the accretion regime will change. Some authors support the blazar sequence \citep[e.g.,][]{Cavaliere2002, Maraschi2003, Maraschi2008, Celotti2008, Ghisellini2008, Ghisellini2010, Chen2011, Finke2013, Ghisellini2010, Ghisellini2017, Chen2021}. Alternatively, \cite{Potter2015} explained the sequence in their inhomogeneous jet model, which is due to the different location of the dissipation region: BL Lacs is closer to the central engine than FSRQs, resulting in the stronger magnetic field, less Compton dominance, and larger peak frequencies concerning FSRQs. 

Since its birth in 1998, the blazar sequence has been an active debate. Some authors believe that this sequence is not true, it is only the result of observation biases \citep[e.g.,][]{Giommi1999, Padovani2003, Nieppola2006, Nieppola2008, Giommi2012, Keenan2021}. \cite{Nieppola2008} explained the observed blazar sequence as a sequence of Doppler factors, that
boosts the intrinsic luminosity and the intrinsic peak frequencies. Considering that the method of deriving the beaming factor uses the  brightness temperature compared to the equipartition
value of $5\times10^{10}$K and the minimum variability timescale observed in the radio. This implies that sources that are not varying during the observation period must have very small Doppler factors, which in fact range from 1 to 30. Therefore, some sources may be under-corrected. It is also considered that the beaming in the radio band may be different from the beaming of more compact and innermost regions producing the bulk of the emission \citep{Prandini2022}. The existence of blazar sequence was, and still is, a very controversial issue. The main objection is that this may be due to the selection effect, which not only worked when it was proposed but also works now, although there are more sensitive instruments and more complete samples, making it reach a deeper sensitivity limit \citep[see also reviews][]{Padovani2007, Ghisellini2008}. To prove this, someone recently proposed an alternative scheme, namely "simplified blazar scenario" \citep{Giommi2012}. \cite{Giommi2012} put forward detailed suggestions on the authenticity of this sequence. They put forward what they called a "simplified solution" for blazars. According to these authors, the blazar sequence is the result of the selection effect, and there is no physical connection between luminosity and the SEDs of blazars. We have both red and blue blazars at high and low luminosities. In this case, it is assumed that the distribution of the $\gamma_{\rm peak}$ reaches the peak at $\sim10^3$. Moreover, it is asymmetric, and the high-energy tail is wider than the low-energy tail. In addition, the radio luminosity function is $N(L)\propto L^{-3}$. It is assumed that the magnetic field of all blazar is $B=0.15$ G, and the median Doppler factor is $<\delta> = 15\pm2$. As a direct result,  there are many low-power BL Lacs with ``red'' SED. These sources have $\gamma_{\rm peak}\sim10^{3}$ or smaller. Nearly 20 years after the publication of the blazar sequence paper, about 1500 blazars were detected by the Fermi-LAT Telescope, as reported by the 3LAC catalog \cite{Ackermann2015}, \cite{Ghisellini2017} proposed a revised sequence by using 747 sources, named Fermi blazar sequence. This time, the authors explored the existence of SED sequences on a sample of gamma-ray selected blazars binned according to the luminosity in the MeV-GeV energy range. \cite{Ghisellini2017} found a significant inverse correlation between synchrotron peak and gamma-ray luminosity by using the third catalog of AGN detected by Fermi/LAT and supported the original blazar sequence. Although a large number of sources have been considered, two types of sources may not have been studied. These are MeV blazars and Extreme TeV blazars.

\cite{Keenan2021} mainly used 3FGL (Fermi Large Area Telescope Third Source Catalogue), Low-frequency radio, and X-ray band data to select jetted AGNs, and found that there was no blazar sequence. However, we have noticed that the synchrotron peak frequency and synchrotron peak frequency luminosity are estimated by using the radio core luminosity, the radio–optical spectral index ($\alpha_{r,o}$), and the optical–X-ray spectral index ($\alpha_{o,x}$) in the sample of \cite{Keenan2021}. The radio galaxies have a weaker beaming effect than the blazar. Therefore, the estimation of synchrotron peak frequency luminosity using radio core luminosity may lead to an underestimate of synchrotron peak frequency luminosity, especially at low peak frequency. To sum up, this may be the reason why they did not find the inverse  correlation between synchrotron peak frequency and synchrotron peak frequency luminosity. At the same time, we find that most of the samples are mainly selected based on Low-frequency radio and/or X-ray emssions. Some authors have found that the Low-frequency radio radiation of some sources comes from star formation \cite[e.g,][]{Caccianiga2015, Jarvela2022}. Therefore, it may not be very accurate to select the jetted AGNs by using the Low-frequency radio band. One question, however, is how to choose a real jetted AGNs. Since the successful launch of the Fermi-Large Area Telescope (Fermi-LAT; \cite{Atwood2009}), many sources have been detected with gamma-ray radiation, including blazars, radio galaxies, and NLS1s. This result shows that these Fermi sources have strong relativistic jets, especially the NLS1s. Therefore, gamma-ray luminosity is a useful tool for selecting real jetted AGNs.  

In the process of studying the SEDs of blazars, in addition to the synchrotron peak frequency and synchrotron peak frequency luminosity, the energy spectrum curvature is also an important physical parameter. If a log-parabolic law is adapted to fit the peaked
component, i.e., $\log\nu f_{\rm \nu}=-b(\log\nu-\log\nu_{\rm p})^{2}+\log\nu_{\rm p}f_{\rm \nu_{\rm p}}$, the
b is the curvature of the SED peak \citep{Chen2014}. Some authors found a relation between the synchrotron peak frequency and the curvature \citep[e.g.,][]{Massaro2004, Massaro2006, Tramacere2007, Tramacere2009}. According to previous theoretical studies, the relationship between peak frequency and curvature of synchrotron can be explained within the framework of the acceleration process of electron emission \cite[see][]{Massaro2006, Tramacere2007, Paggi2009, Chen2014}. \cite{Chen2014} theoretically analyzed the relationship between the peak frequency and the curvature of the synchrotron, and proposed two electron acceleration mechanisms according to the relationship coefficient between the peak frequency and the curvature of the synchrotron. \cite{Chen2014} proposed that for the  stochastic acceleration $k$ ($1/b_{\rm sy}=k\log\nu_{\rm p}+a$), the slope of the relationship between the peak frequency and the curvature of the synchrotron is $k=2$, for statistical acceleration with energy-dependent acceleration is 2.5, and for fluctuations of fractional acceleration gain is 3.33. \cite{Chen2014} found that the slope of this relation was k = 2 by using 48 fermi blazars, which is consistent with stochastic acceleration.       

With the release of the fourth source catalog data of Fermi Large Area Telescope (LAT), we can use a large sample of Fermi sources to re-examine blazar phenomenology and in particular the evidence for Fermi blazar sequence and the particle acceleration mechanism of jetted AGNs. One of the motives of this work is to check the more sensitive instruments we now infer: Fermi/LAT is 20 times more sensitive than EGRET, and patrols the whole sky in a more effective way. Gamma-ray is usually the main part of the electromagnetic output. Because of the abundance of gamma-ray sources, we use Fermi detection and redshift knowledge as the only selection criteria. This is different from the original blazar sequence, in which objects are selected based on their radio and/or X-ray emission.
Moreover, we can select the real jetted AGNs by using $\gamma$-ray. At present, it is unclear whether the $\gamma$NLS1s and radio galaxies belong to the Fermi blazar sequence. Therefore, we use a large sample to study the relationship between the Fermi blazar sequence, radio galaxies, and $\gamma$NLS1s. At the same time, we studied the particle acceleration mechanism for these Fermi sources based on the curvature–synchrotron peak frequency relation. The paper is structured as follows. In Section 2 we present the sample; in Section 3 we present the results and discussion; Section 4 describes the conclusions. A $\Lambda$CDM cosmology with $H_{0}=70 \rm km~s^{-1}Mpc^{-1}$, $\Omega_{\rm m}=0.27$, $\Omega_{\Lambda}=0.73$ is adopted. 

\section{The sample}
\subsection{The fermi blazar and $\gamma$NLS1s sample}    
We try to collect large samples of Fermi sources with reliable redshift, black hole mass, accretion disk luminosity, synchrotron peak frequency, and synchrotron peak frequency luminosity. We consider the sample of \cite{Paliya2021}. \cite{Paliya2021} got 1077 sources with black hole mass and disk luminosity. A detailed description of the calculation of black hole mass and the luminosity of accretion disk is presented in the work of \cite{Paliya2021}. \cite{Paliya2021} used a second-degree
polynomial to fit quasi-simultaneous multiwavelength data of these fermi sources using the built-in functionality provided in the SSDC SED builder tool \footnote{ https://tools.ssdc.asi.it/SED/} and got the synchrotron and inverse Compton peak frequencies and peak frequencies flux. We use the ratio of the high- and low-energy peak luminosities to obtain the Compton Dominance (CD). We carefully checked the sample of \cite{Paliya2021} and compared it with the source classification of \cite{Abdollahi2020} and \cite{Foschini2021}, and found that 17 $\gamma$NLS1s were included in the sample of \cite{Paliya2021}. We only consider these sources with 1.4 GHz radio flux from the NED. Finally, we get 504 FSRQs, 277 BL Lacs, and 17 $\gamma$NLS1s. The data is listed in Table 1.    

\cite{Komossa2018} used the following formula to estimate the jet kinetic power of AGNs \citep{Birzan2008}, 
\begin{eqnarray}
	\log P_{\rm jet}=0.35(\pm0.07)\log P_{1.4} + 1.85(\pm 0.10)
\end{eqnarray}
where $P_{\rm jet}$ is in units $10^{42} \rm erg~s^{-1}$, and $P_{1.4}$ in units $10^{40} \rm erg~s^{-1}$. The $P_{1.4}$ is 1.4 GHz radio luminosity,  $P_{\rm 1.4}=4\pi d_{\rm L}^{2}\nu S_{\rm \nu}$, $S_{\rm \nu}= S_{\rm \nu}^{\rm obs}(1+z)^{\alpha-1}$, where $\alpha$ is spectral index, $\alpha=0$ is assumed \citep{Donato2001, Abdo2010, Komossa2018}. \cite{Komossa2018} believes that if the relationship of \cite{Cavagnolo2010} is used instead, the predicted value of $P_{\rm jet}$ can be up to an order of magnitude. Thus, we estimate the jet kinetic power of these fermi sources by using equation (1). \cite{Foschini2015} used the 15 GHz radio core luminosity to estimate the jet kinetic power of F-RLNLS1s, $\log P_{\rm jet, kinetic}=(0.90\pm0.04)\log L_{\rm radio, core}+(6\pm2)$. We compare the jet kinetic power calculated using Equation (1) with that calculated using Equation of \cite{Foschini2015}. The jet kinetic powers calculated using Equation (1) are $\log P_{\rm jet, kinetic}=44.32$ (1H 0323+342), $\log P_{\rm jet, kinetic}=45.03$ (SBS 0846+513), $\log P_{\rm jet, kinetic}=44.83$ (PMN J0948+0022), $\log P_{\rm jet, kinetic}=44.93$ (PKS 1502+036), $\log P_{\rm jet, kinetic}=44.40$ (FBQS J1644+2619), and $\log P_{\rm jet, kinetic}=44.88$ (PKS 2004-447). The jet kinetic powers calculated using Equation of \cite{Foschini2015} are $\log P_{\rm jet, kinetic}=43.40$ (1H 0323+342), $\log P_{\rm jet, kinetic}=45.09$ (SBS 0846+513), $\log P_{\rm jet, kinetic}=45.27$ (PMN J0948+0022), $\log P_{\rm jet, kinetic}=45.08$ (PKS 1502+036), $\log P_{\rm jet, kinetic}=43.57$ (FBQS J1644+2619), and $\log P_{\rm jet}=44.36$ (PKS 2004-447). We do not consider the radiation part, but only the kinetic energy part when calculating the jet power. We find that the largest difference between them is less than one order of magnitude. 

\subsection{The radio galaxies sample} 
We consider the radio galaxies with 1.4 GHz radio flux from the 4FGL-DR2 catalog and select 38 samples. The 1.4 GHz radio flux comes from the NED. The black hole mass of radio galaxies is estimated by using the following formula \citep{Marconi2003, Sbarrato2014},

\begin{equation}
	\log\left(\frac{M}{M_{\rm \odot}}\right) = -2.77-0.464\times M_{\rm H}
\end{equation} 
where $M_{\rm H}$ is the absolute magnitude in the $H$ band. We get $M_{\rm H}$ from the NED and use equation (2) to estimate the black hole mass of radio galaxies. \cite{Buttiglione2010} got the disk luminosity of 3CR sample. We cross-match with their sample. We only get the disk luminosity of seven radio galaxies. For the remaining 31 radio galaxies, we use the synchrotron peak frequency luminosity to estimate the luminosity of the accretion disk, i.e, $L_{\rm disk}\sim 2\nu L_{\rm \nu}$ \citep{Calderone2013, Ghisellini2015}. It should be noted that the luminosity of the accretion disk obtained by this method is slightly larger than that obtained by optical spectra (see Figure 1). We also use equation (1) to estimate the jet kinetic power of radio galaxies. We also collect quasi-simultaneous multiwavelength data of radio galaxies from the Space Science Data Center (SSDC) SED builder tool. We use a second-degree polynomial to fit the quasi-simultaneous multiwavelength data of these radio galaxies and get the synchrotron and inverse Compton peak frequencies and corresponding flux values.    

\begin{table*}
	\caption{The sample of jetted AGN}
	\centering
	\label{table1}
	\setlength{\tabcolsep}{1.0mm}{
		\begin{tabular}{llllllllccccccrrrrr} % four columns, alignment for each
			\hline
			\hline
			Name & Type & Redshift & $log(M/M_{\odot})$  & $logL_{disk}$ & $logL_{\gamma}$ &  $f_{1.4~GHz}$   & log$P_{jet}$ & $log\nu_{pk}^{sy}$ & $CD$ & $b_{sy}$ & $logL_{pk}^{sy}$ \\
			(1)    &  (2)  & (3)   & (4)  &  (5) & (6) & (7) & (8) & (9) & (10) & (11) & (12)\\
			\hline
			J0001.5+2113	&	FSRQ	&	0.439	&	7.539 	&	44.65 	&	46.54	&	0.217	&	44.87 	&	13.81	&	30.9	&	0.117 	&	45.17 	\\
			J0003.2+2207	&	BLL	&	0.1	&	8.100 	&	42.74 	&	43.71	&	0.0087	&	43.84 	&	15.15	&	0.21	&	0.102 	&	43.23 	\\
			J0004.4-4737	&	FSRQ	&	0.88	&	8.280 	&	45.10 	&	47.16	&	0.932	&	45.43 	&	13.01	&	2.4	&	0.110 	&	46.61 	\\
			J0006.3-0620	&	BLL	&	0.347	&	8.924 	&	44.52 	&	45.01	&	2.051	&	45.12 	&	12.92	&	0.11	&	0.180 	&	45.74 	\\
			J0010.6+2043	&	FSRQ	&	0.598	&	7.861 	&	45.34 	&	45.94	&	0.158	&	44.96 	&	12.42	&	1.41	&	0.237 	&	45.47 	\\
			J0011.4+0057	&	FSRQ	&	1.491	&	8.664 	&	45.71 	&	48.2	&	0.167	&	45.51 	&	12.81	&	3.63	&	0.196 	&	47.30 	\\
			J0013.6+4051	&	FSRQ	&	0.256	&	7.022 	&	43.13 	&	44.8	&	1.65	&	44.97 	&	12.73	&	0.46	&	0.120 	&	44.99 	\\
			J0013.6-0424	&	FSRQ	&	1.076	&	7.816 	&	45.03 	&	47.07	&	0.304	&	45.38 	&	12.3	&	1.55	&	0.220 	&	46.42 	\\
			J0013.9-1854	&	BLL	&	0.095	&	9.650 	&	43.27 	&	43.91	&	0.0295	&	44.01 	&	17.43	&	0.08	&	0.058 	&	44.02 	\\
			J0014.1+1910	&	BLL	&	0.477	&	7.465 	&	44.32 	&	45.66	&	0.154	&	44.86 	&	13.74	&	1.41	&	0.136 	&	45.46 	\\
			J0014.2+0854	&	BLL	&	0.163	&	8.850 	&	43.37 	&	44.42	&	0.326	&	44.55 	&	15.64	&	0.43	&	0.051 	&	43.75 	\\
			J0014.3-0500	&	FSRQ	&	0.791	&	7.928 	&	44.93 	&	46.83	&	0.0318	&	44.86 	&	12.6	&	2.75	&	0.277 	&	45.98 	\\
			J0015.6+5551	&	BLL	&	0.217	&	9.680 	&	44.05 	&	44.87	&	0.0849	&	44.45 	&	17.11	&	0.35	&	0.052 	&	44.67 	\\
			J0016.2-0016	&	FSRQ	&	1.577	&	8.522 	&	45.77 	&	48.73	&	0.957	&	45.81 	&	12.44	&	10.47	&	0.153 	&	47.22 	\\
			J0016.5+1702	&	FSRQ	&	1.721	&	8.874 	&	45.74 	&	48.41	&	0.135	&	45.58 	&	12.79	&	4.79	&	0.184 	&	47.37 	\\
			J0017.5-0514	&	FSRQ	&	0.227	&	7.831 	&	44.74 	&	45.46	&	0.178	&	44.58 	&	12.61	&	4.9	&	0.259 	&	44.83 	\\
			J0017.8+1455	&	BLL	&	0.303	&	8.270 	&	44.16 	&	45.13	&	0.0595	&	44.52 	&	14.22	&	0.49	&	0.126 	&	44.84 	\\
			J0019.6+7327	&	FSRQ	&	1.781	&	9.306 	&	46.62 	&	49.32	&	1.25	&	45.95 	&	12.29	&	14.12	&	0.182 	&	47.77 	\\
			J0021.6-0855	&	BLL	&	0.648	&	8.540 	&	44.63 	&	46.06	&	0.0472	&	44.82 	&	14.22	&	1.26	&	0.109 	&	45.33 	\\
			J0022.0+0006	&	BLL	&	0.306	&	8.020 	&	43.79 	&	44.9	&	0.0042	&	44.13 	&	16.67	&	0.31	&	0.073 	&	44.57 	\\
			J0023.7+4457	&	FSRQ	&	1.062	&	7.709 	&	45.09 	&	47.49	&	0.141	&	45.26 	&	12.8	&	6.03	&	0.202 	&	46.56 	\\
			J0024.7+0349	&	FSRQ	&	0.546	&	7.114 	&	44.62 	&	45.91	&	0.022	&	44.62 	&	13.47	&	2.69	&	0.172 	&	45.15 	\\
			J0025.2-2231	&	FSRQ	&	0.834	&	8.492 	&	45.71 	&	46.41	&	0.202	&	45.17 	&	13.97	&	2.63	&	0.092 	&	45.71 	\\
			J0028.4+2001	&	FSRQ	&	1.553	&	8.426 	&	45.57 	&	48.37	&	0.287	&	45.62 	&	12.29	&	2.57	&	0.239 	&	47.36 	\\
			J0032.4-2849	&	BLL	&	0.324	&	8.470 	&	44.02 	&	45.16	&	0.161	&	44.70 	&	13.78	&	1.26	&	0.133 	&	44.95 	\\
			J0038.2-2459	&	FSRQ	&	0.498	&	8.139 	&	44.97 	&	45.92	&	0.413	&	45.03 	&	12.43	&	1.86	&	0.233 	&	45.59 	\\
			J0039.0-0946	&	FSRQ	&	2.106	&	8.499 	&	45.73 	&	49.19	&	0.154	&	45.76 	&	12.53	&	9.12	&	0.183 	&	47.59 	\\
			J0040.4-2340	&	BLL	&	0.213	&	8.680 	&	43.75 	&	44.62	&	0.0536	&	44.38 	&	14.44	&	0.55	&	0.102 	&	44.00 	\\
			J0040.9+3203	&	FSRQ	&	0.632	&	7.173 	&	44.47 	&	46.09	&	0.414	&	45.14 	&	13.14	&	1.91	&	0.109 	&	45.17 	\\
			J0042.2+2319	&	FSRQ	&	1.425	&	8.733 	&	45.49 	&	48.02	&	1.27	&	45.78 	&	12.27	&	5.25	&	0.151 	&	46.92 	\\
			J0043.8+3425	&	FSRQ	&	0.969	&	7.828 	&	44.76 	&	47.81	&	0.0933	&	45.14 	&	13.77	&	12.3	&	0.111 	&	45.89 	\\
			\hline
	\end{tabular}}
	\tablecomments{Columns (1) is the name of sources; Columns (2) is the type of sources; Columns (3) is redshift; Columns (4) is the black hole mass; Columns (5) is the disk luminosity, units is erg$~s^{-1}$; Columns (6) is the observation gamma-ray luminosity, units is erg$~s^{-1}$; Columns (7) is the 1.4 GHz radio flux in units jy; Columns (8) is the jet kinetic power in units erg$~s^{-1}$; Columns (9) is the synchrotron peak frequency; Columns (10) is the Compton dominance; Columns (11) is the curvature; Columns (12) is the synchrotron peak frequency luminosity. This table is available in its entirety in machine-readable form.}
\end{table*}

\section{Result and Discussion}
\subsection{The Fermi blazar sequence}
Figure 2 shows the relation between synchrotron peak luminosity and peak frequency. We find an ``L'' shape in this figure. \cite{Nieppola2006} found an ``V'' shape by using BL Lacs, excluding FSRQ; If the FSRQ with high luminosity and low peak is added, it may look more like an "L" shape. \cite{Meyer2011} also found an ``L'' shape by using 216 blazars and 41 radio galaxies. We confirm their results. We also note that there is no inverse correlation between synchrotron peak luminosity and peak frequency for most of the sources with $\nu_{peak}>10^{15}$Hz. We carefully examine these sources with a peak frequency greater than $10^{15}$Hz and found 176 BL Lacs. According to the TeVCat catalogue\footnote{http://tevcat2.uchicago.edu/}, we find that 19 of the 176 sources are TeV sources, about 10.8 percent. Extreme TeV blazars are expected to have a steady but faint emissions in the GeV-TeV range. As a result, the fainter or more distant objects might easily remain below the sensitivity limit of Fermi. These subclasses represent the extremes of the sequence, and it is very important to test the current understanding of the blazar sequence through future experiments in the GeV-TeV range. Another key feature of this class of source is its weak luminosity. For known extreme TeV blazars, the measured luminosity is only slightly higher than the Fermi LAT sensitivity limit in the gamma-ray range. In the context of the Fermi blazar sequence, these extreme TeV blazars that are still missing would
populate the lowest luminosity range. At present, due to the small number of TeV samples, we can not infer whether the TeV source causes the peak frequency and peak frequency luminosity to have no anti-correlation when the peak frequency is greater than $10^{15}$Hz. Current  gamma-ray instruments like the Large High Altitude Air Shower Observatory (LHAASO), future Cherenkov Telescope Array Observatory (CTAO), and so on, will be the key to probing this expected trend. 

From Figure 2, we also find that the synchrotron peak frequency of $\gamma$NLS1s lies below $10^{14}$ Hz, which suggests that these $\gamma$NLS1s are low-synchrotron-peaked (LSP) type beamed AGNs. The range of synchrotron peak frequency of most radio galaxies is from $10^{14}$ to $10^{16}$~Hz, which suggests that these radio galaxies are intermediate-synchrotron-peaked (ISP) AGNs. The $\gamma$NLS1s and radio galaxies follow a fermi blazar sequence of the ``L'' shape, which implies that $\gamma$NLS1s and radio galaxies belong to the fermi blazar sequence. In the future, due to the improvement of the sensitivity of the instrument, many sources may also be detected to have TeV radiation, so the blazar sequence can be well tested by using large samples.          

\begin{figure}
    \includegraphics[width=8.5cm,height=8.5cm]{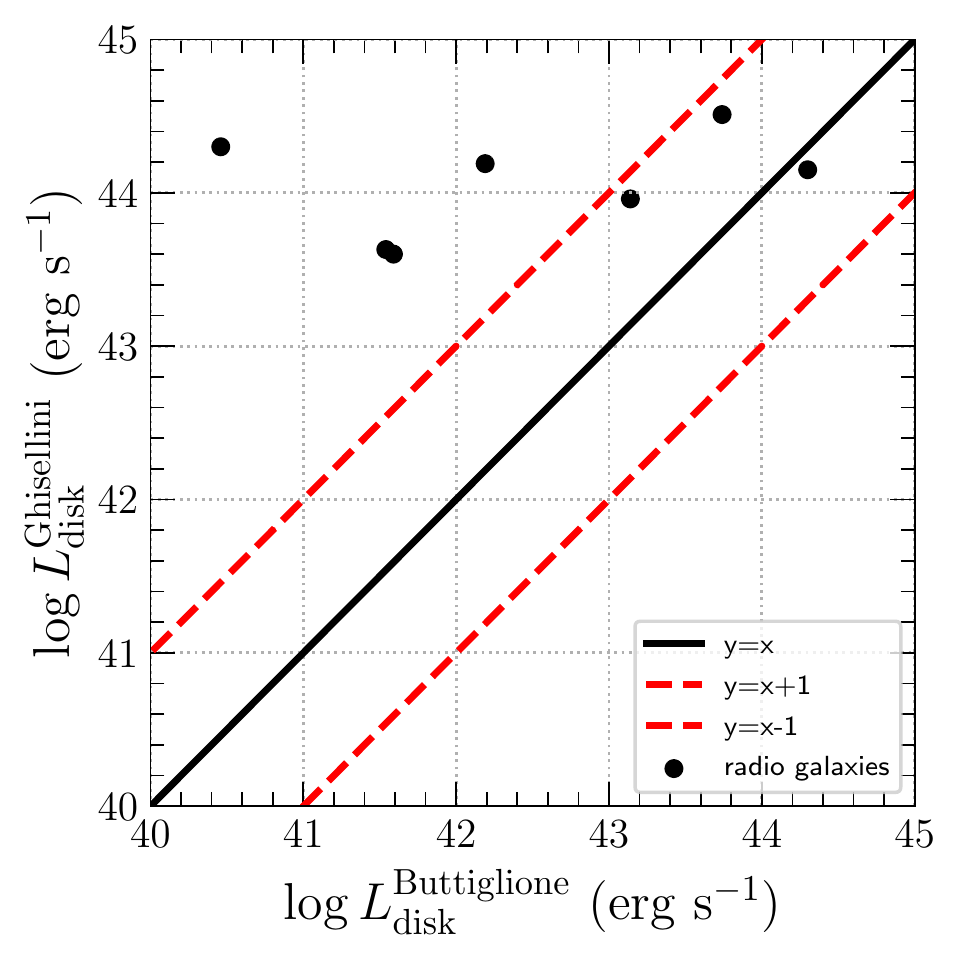}
    \caption{The accretion disk luminosity of \cite{Ghisellini2015} vs. the disk luminosity of \cite{Buttiglione2010}. The black dot is radio galaxies.The y = x an equal line. The red dashed lines y = x$\pm$1 give an idea of one order of magnitude difference with respect to the 1:1 ratio.}
    \label{Figure}
\end{figure}

\begin{figure}
	\includegraphics[width=8.5cm,height=8.5cm]{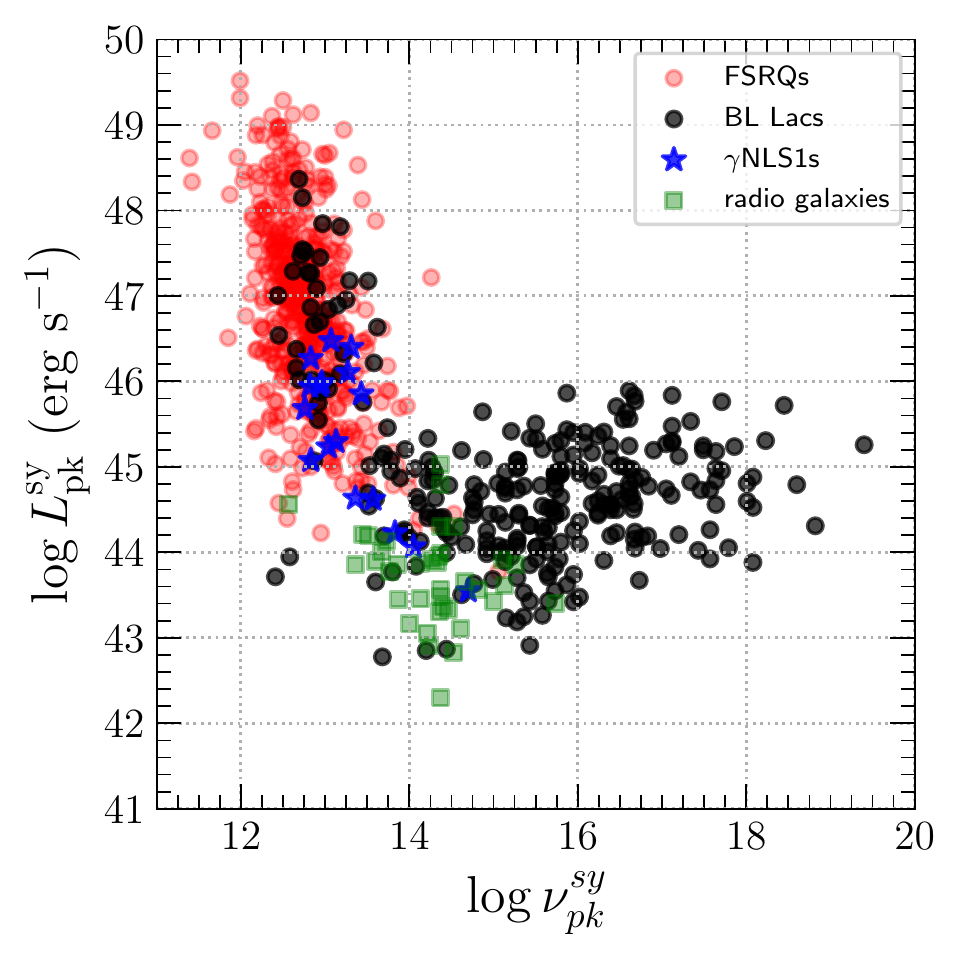}
	\caption{Relation between synchrotron peak luminosity ($L_{\rm pk}^{\rm sy}$) and peak frequency ($\nu_{\rm pk}^{\rm sy}$) for the whole sample. FSRQs are shown as
		red-filled circles, BL Lacs as black-filled circles, $\gamma$NLS1s as stars, and radio galaxies as squares. The black line is the equal line.}
	\label{figure1}
\end{figure}

In addition, it is found that there is a significant inverse correlation between the 5 GHz luminosity and the peak frequency of synchrotron \citep{Fossati1998}. The 5 GHz luminosity is an indication of jet power. Therefore, we study the relationship between jet kinetic power, $\gamma$-ray luminosity, and synchrotron peak frequency. In the top panel of Figure 3, we show the relation between jet kinetic power, $\gamma$-ray luminosity, and synchrotron peak frequency, respectively. We find that there is no anti-correlation between jet kinetic power, $\gamma$-ray luminosity, and synchrotron peak frequency for most of BL Lacs with high synchrotron peak frequency ($\nu>10^{15}$Hz). \cite{Keenan2021} also found a similar tendency. They define these sources with $\nu>10^{15}$Hz as sources of weak/type I jets. These sources with weak jets are inefficient accretion. We also find that these sources with weak jets have low $\log L_{disk}/L_{Edd}$, which suggests that these sources are inefficient accretion. Our results are consistent with the work of \cite{Keenan2021}. The radio galaxies have lower $\gamma$-ray luminosity than other type AGNs, which suggests that the radio galaxies have a weak beaming effect. \cite{Meyer2011} found that Fermi radio galaxies have low Core dominance parameter ($R_{CE}$) than FSRQ and BL Lacs, which confirmed that these Fermi radio galaxies have a weak beaming effect.        

The $\gamma$-ray dominance was discussed as part of the original ``blazar sequence'' by \cite{Fossati1998}. Some authors have studied the relation between $\gamma$-ray dominance or Compton dominance and synchrotron peak frequency, in favor of testing the correlation between $L_{\rm pk}^{\rm sy}$ and $\nu_{\rm pk}^{\rm sy}$ \citep[e.g.,][]{Padovani2003, Nieppola2006, Giommi2012, Finke2013}. We also test this correlation. Pearson analysis shows that there is a significant inverse correlation between Compton dominance and synchrotron peak frequency for the whole sample ($r=-0.69$, $P=5.66\times10^{-121}$). Spearman and Kendall tau are used to detect this correlation. The Spearman correlation coefficient and significance level are $r=-0.67$ and $P=4.13\times10^{-112}$. The Kendall tau correlation coefficient and significance level are $r=-0.48$ and $P=1.75\times10^{-94}$. The fitted equation is as follows,

\begin{equation}
	\log CD=(-0.29\pm0.01)\log\nu_{\rm pk}^{\rm sy}+(4.13\pm0.14).
\end{equation}    

The validity of blazar sequence has been debated for a long time  \citep{Fossati1998}, i.e., whether such a sequence has a
physical origin \citep{Ghisellini1998}. From the right bottom of figure 3, we also find that FSRQs and $\gamma$NLS1s have high accretion rates and low synchrotron peak frequency, while BL Lacs and radio galaxies have low accretion rates and high synchrotron peak frequency. \cite{Ghisellini2008} have suggested that the jet power and SED of blazars are mainly related to two physical parameters of the accretion process, namely the black hole mass and the accretion rate. \cite{Ghisellini1998} suggested that radiative cooling leads to the observational phenomenon of the blazar sequence. FSRQ and $\gamma$NLS1 have high accretion rate, which leads to fast cooling of relativistic electrons \citep[see also][]{Foschini2017}. FSRQ and $\gamma $NLS1 have lower synchrotron peak frequencies. However, the accretion rate of BL Lac and radio galaxies is very low, which leads to the slow cooling of relativistic electrons. BL Lacs and radio galaxies have high synchrotron peak frequencies. The Pearson analysis shows a significant anti-correlation between accretion rate and synchrotron peak frequency for the whole sample ($r=-0.74$, $P=3.8\times10^{-144}$). The Spearman correlation coefficient and significance level are $r=-0.70$ and $P=3.13\times10^{-122}$. The Kendall tau correlation coefficient and significance level are $r=-0.49$ and $P=5.92\times10^{-98}$. The fitted equation is as follows,

\begin{equation}
	\log\frac{L_{\rm disk}}{L_{\rm Edd}}=(-0.57\pm0.02)\log\nu_{\rm pk}^{\rm sy}+(6.19\pm0.25).
\end{equation}    
         
\begin{figure*}
	\includegraphics[width=16cm,height=16cm]{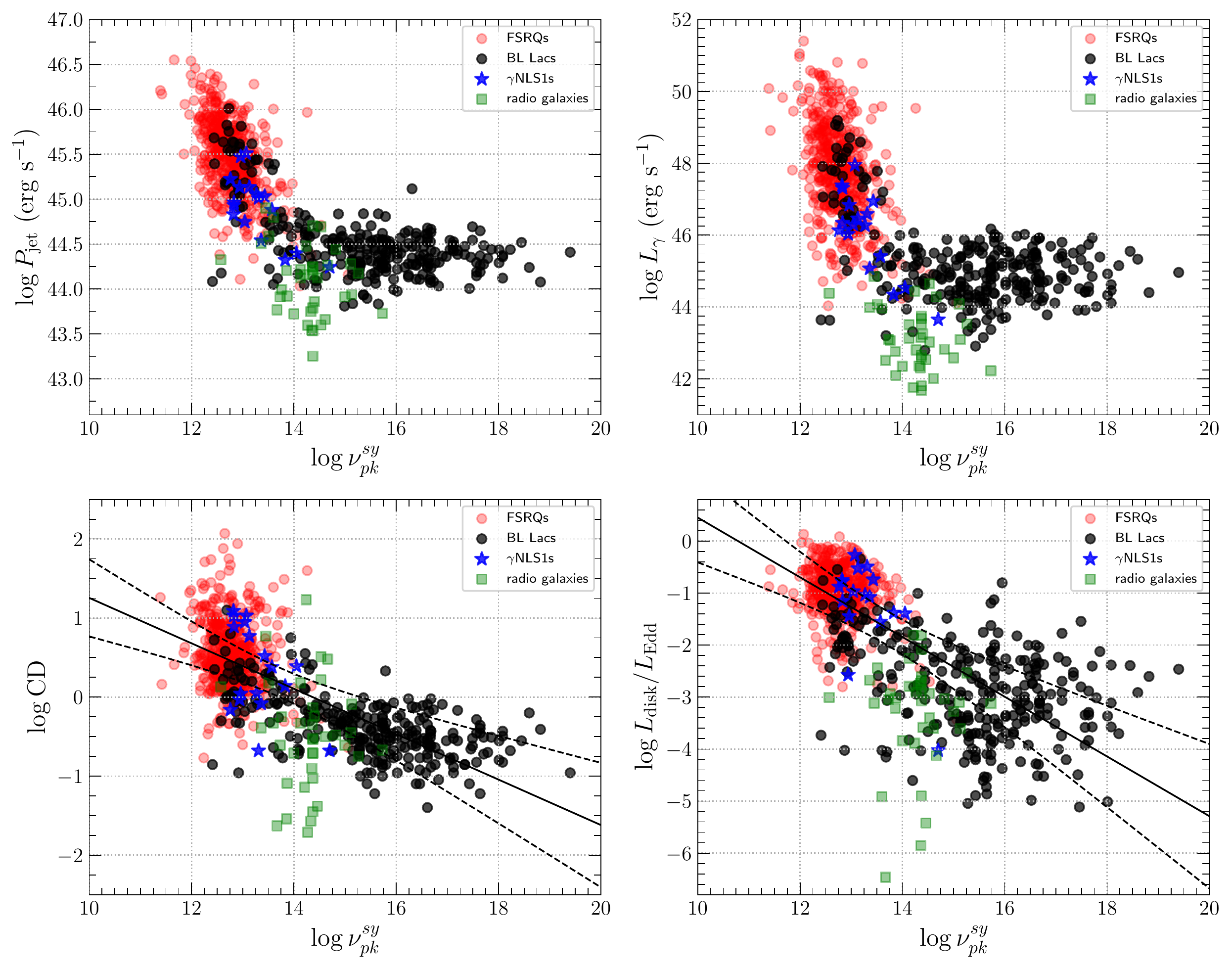}
	\centering
	\caption{Top: relation between the jet kinetic power (left), $\gamma$-ray luminosity (right) and synchrotron peak frequency. Bottom: the Compton dominance (left) and the accretion luminosity (right, in Eddington units) versus synchrotron peak frequency. The solid line corresponds to the best fitting linear model obtained by symmetric least square fitting. The dotted line represents the confidence range of $3\sigma$.}
	\label{figure2}
\end{figure*}

In the physical scheme, a luminous disk implies an efficient accretion process
($L_{\rm disk}/L_{\rm Edd}\geq1\times10^{-2}$), which ionizes the broad-line region clouds, thus detecting the broad optical emission lines. Therefore, the jet electrons interact with the dense photon field in broad-line region through the EC (External Compton) mechanism, and then reach high energy to produce luminous gamma-ray emission, which makes the SED more dominated by Compton and reaches its peak at low frequency. On the other hand, in the
low-accretion regime ($L_{\rm disk}/L_{\rm Edd}<1\times10^{-2}$), the external
radiation field becomes weaker and the SED is less Compton dominated. We test these hypotheses on our sample of jetted AGNs, and the results are shown in Figure 4.

\begin{figure*}
	\includegraphics[width=16cm,height=16cm]{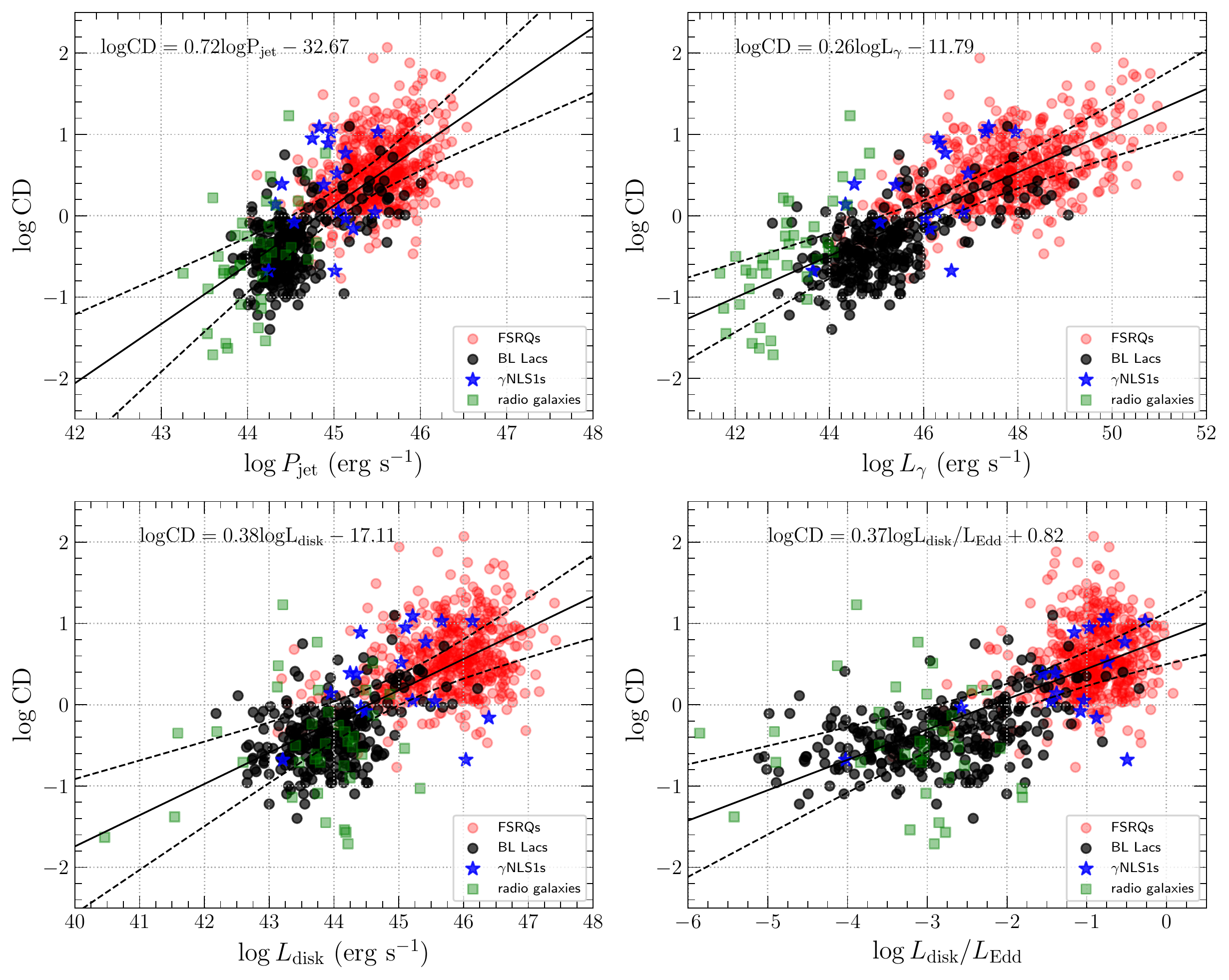}
	\centering
	\caption{Top: the Compton dominance versus jet kinetic power (left) and $\gamma$-ray luminosity (right). Bottom: the Compton dominance versus the accretion disk luminosity (left) and the accretion disk luminosity (right, in Eddington units). The solid and dashed lines are the same as in figure 3.}
	\label{figure3}
\end{figure*}

In the top panels of Figure 4, we show the relation between Compton dominance and jet kinetic power and $\gamma$-ray luminosity, respectively. Strong positive correlations are found in both cases. The Pearson analysis shows a significant correlation between Compton dominance and jet kinetic power ($r=0.74$, $P=6.34\times10^{-147}$) and $\gamma$-ray luminosity ($r=0.77$, $P=1.81\times10^{-170}$) for the whole sample, respectively. The Spearman correlation coefficient and significance level for the relation between Compton dominance and jet kinetic power are $r=0.73$ and $P=1.36\times10^{-141}$, for the relation between Compton dominance and $\gamma$-ray luminosity are $r=0.77$ and $P=5.05\times10^{-169}$. The Kendall tau correlation coefficient and significance level for the relation between Compton dominance and jet kinetic power are $r=0.53$ and $P=3.51\times10^{-115}$, for the relation between Compton dominance and $\gamma$-ray luminosity are $r=0.57$ and $P=9.41\times10^{-135}$. We find that FSRQs and $\gamma$NLS1s with high $\gamma$-ray luminosity are Compton dominance, which implies that the high-energy $\gamma$-ray emission is mainly provided by the external Compton mechanism \citep[e.g.,][]{Sikora1994, Ghisellini2008, Ghisellini2010, Finke2013, Paliya2019}. However, the high-energy component of the BL Lacs and radio galaxies is dominated by the synchrotron self-Compton (SSC) process \citep[e.g.,][]{Ghisellini2010, Lister2011, Ackermann2012, Zhang2013}. 

From figure 4, we get the average values of jet kinetic power and Compton dominance for FSRQs are $\log P_{\rm jet}=45.48\pm0.41$ and $\log CD=0.50\pm0.43$, for BL Lacs are $\log P_{\rm jet}=44.44\pm0.40$ and $\log CD=-0.41\pm0.39$, for $\gamma$NLS1s are $\log P_{\rm jet}=44.96\pm0.36$ and $\log CD=0.38\pm0.56$, and for radio galaxies are $\log P_{\rm jet}=44.14\pm0.38$ and $\log CD=-0.51\pm0.67$. \cite{Foschini2015} got that the average values of jet power of F-RLNLS1s are $\log P_{\rm kin}=43.62$. We find that there is no more than one order of magnitude difference between our results and theirs within the error range. Our results are consistent with those of \cite{Bottcher2002} and \cite{Cavaliere2002}, implying the evolution of high-power, Compton-dominated sources to less-luminous,
less Compton-dominated ones. At the same time, the evolutionary sequence of blazar shows that quasars are young AGNs. With their growth, they become BLLac objects \citep{Cavaliere2002, Bottcher2002, Maraschi2003}. NLS1 is considered to be a low redshift analogue of early quasars \citep{Mathur2000}. \cite{Berton2016} clearly shows that NLS1s might be the young counterpart of FSRQs \citep[see also][]{Foschini2015}. These results may imply that the FSRQs, BL Lacs, $\gamma$NLS1s, and radio galaxies seem to form an evolutionary sequence, $\gamma$NLS1s$\rightarrow$FSRQs$\rightarrow$BL Lacs$\rightarrow$radio galaxies. However, we warn that strong claims cannot be made for the following reasons. In our sample, there are very few sources exceeding redshift 3. In our future work, we will try to
address these shortcomings by considering $\gamma$-ray  detected high-redshift sources.       

The relations between Compton dominance and accretion disk luminosity (left) and accretion disk luminosity in Eddington units (right) are shown in the bottom panels of Figure 4. The Pearson analysis shows a significant correlation between Compton dominance and disk luminosity ($r=0.69$, $P=5.01\times10^{-123}$) and disk luminosity in Eddington units ($r=0.70$, $P=4.34\times10^{-125}$) for the whole sample, respectively. The Spearman correlation coefficient and significance level for the relation between Compton dominance and disk luminosity are $r=0.69$ and $P=2.55\times10^{-123}$, for the relation between Compton dominance and disk luminosity in Eddington units are $r=0.70$ and $P=2.22\times10^{-122}$. The Kendall tau correlation coefficient and significance level for relation between Compton dominance and disk luminosity are $r=0.49$ and $P=7.7\times10^{-101}$, for the relation between Compton dominance and disk luminosity in Eddington units are $r=0.49$ and $P=3.39\times10^{-100}$. The empirical relation between Compton dominance and disk luminosity in Eddington units is as follows,

\begin{equation}
	\log CD=(0.37\pm0.01)\log\frac{L_{\rm disk}}{L_{\rm Edd}}+(0.82\pm0.03).
\end{equation}     

From the bottom left panel of figure 4, we can see that FSRQs and $\gamma$NLS1s with high disk luminosity tend to have high Compton dominance, which hints that more Compton-dominated sources host more powerful disks. This anti-correlation also indicates a physical connection between the luminosity of accretion disk and the behavior of the SED in jetted AGNs. Stronger accretion disk emission means a luminous broad-line region (BLR), whose strong radiation field interacts with jet electrons, making them lose energy mainly through EC-BLR process. The bottom right panel of figure 4 shows the obvious difference of two groups in regions defined by $CD>1$, $L_{\rm disk}/L_{\rm Edd}>0.01$ for both FSRQs and $\gamma$NLS1s and $CD<1$, $L_{\rm disk}/L_{\rm Edd}<0.01$ for both BL Lacs and radio galaxies. Based on this finding, we suggest that FSRQs and $\gamma$NLS1s are high-Compton-dominated (HCD) with $CD>1$, and BL Lacs and radio galaxies are low-Compton-dominated (LCD) sources for $CD<1$. At the same time, FSRQs have $CD>1$, $L_{\rm disk}/L_{\rm Edd}>0.01$ and BL Lacs have $CD<1$, $L_{\rm disk}/L_{\rm Edd}<0.01$. This classification scheme is analogous to that based on $L_{\rm BLR}/L_{\rm Edd}$ proposed by \cite{Ghisellini2011}. Based on these findings and the results discussed above, we can conclude that accretion rate plays an important role in controlling the observational characteristics of powerful blazar, which supports the statement that blazar sequence has a physical origin \citep[e.g.,][]{Ghisellini2017}. 

\subsection{Particle acceleration mechanisms}   
The statistical acceleration and the stochastical acceleration mechanisms can be used to explain the correlation between $\log\nu_{\rm pk}^{\rm sy}$ and $1/b_{\rm sy}$ \citep[see details in][]{Chen2014}. \cite{Chen2014} suggested that within the framework of the statistical acceleration, it depends on the acceleration probability of energy or the fluctuation of fractional acceleration gain. For the case of energy-dependent acceleration probability, \cite{Chen2014} got $1/b_{\rm sy}\approx 5/2\log\nu_{\rm pk}^{\rm sy}+C$. For the case of fluctuations of fractional acceleration gain, \cite{Chen2014} got $1/b_{\rm sy}\approx 10/3\log\nu_{\rm pk}^{\rm sy}+C$. The second scenario is in the framework of stochastic acceleration, \cite{Chen2014} got $1/b_{sy}\approx 2\log\nu_{\rm pk}^{\rm sy}+C$.      

Figure 5 shows the relation between $\log\nu_{\rm pk}^{\rm sy}$ and $1/b_{\rm sy}$ for Fermi blazars, $\gamma$NLS1s and radio galaxies. We derive the correlation between $\log\nu_{\rm pk}^{\rm sy}$ and $1/b_{\rm sy}$ for the whole sample ($r=0.89$, $P=1.66\times10^{-287}$)   

\begin{equation}
	1/b_{\rm sy}=(2.62\pm0.05)\log\nu_{\rm pk}^{\rm sy}+(-27.78\pm0.64)
\end{equation}     
and for Fermi blazars ($r=0.92$, $P=1.75\times10^{-310}$)   

\begin{equation}
	1/b_{\rm sy}=(2.63\pm0.04)\log\nu_{\rm pk}^{\rm sy}+(-27.85\pm0.57)
\end{equation}     
and for FSRQs ($r=0.63$, $P=6.48\times10^{-57}$)   

\begin{equation}
	1/b_{\rm sy}=(2.55\pm0.14)\log\nu_{\rm pk}^{\rm sy}+(-26.68\pm1.79)
\end{equation}
and for BL Lacs ($r=0.88$, $P=6.06\times10^{-92}$)   

\begin{equation}
	1/b_{\rm sy}=(2.96\pm0.10)log\nu_{\rm pk}^{\rm sy}+(-33.12\pm1.47)
\end{equation}
and for $\gamma$NLS1s ($r=0.61$, $P=0.009$)   

\begin{equation}
	1/b_{\rm sy}=(2.33\pm0.78)\log\nu_{\rm pk}^{\rm sy}+(-24.18\pm10.36)
\end{equation}
and for radio galaxies ($r=0.37$, $P=0.02$)   

\begin{equation}
	1/b_{\rm sy}=(2.23\pm1.38)\log\nu_{\rm pk}^{\rm sy}+(-22.54\pm19.77).
\end{equation}

The slopes of the relation between synchrotron peak frequency and synchrotron curvature for the whole sample are $k_{\rm whole}=2.62\pm0.05$, for Fermi blazars are $k_{\rm Fermi~blazars}=2.63\pm0.04$, for FSRQs are $k_{\rm FSRQs}=2.55\pm0.14$, and for $\gamma$NLS1s are $k_{\rm \gamma NLS1s}=2.33\pm0.78$. We find that these slopes are close to 2.5, which implies that these AGNs can be explained by statistical acceleration for the case of energy-dependent acceleration probability. \cite{Chen2014} found that the slope of the relation was $2.04\pm0.03$ by using 48 blazars, which is consistent with the stochastic acceleration. Our results are different from theirs, which may be due to the number of samples.

The slope of the relation between synchrotron peak frequency and synchrotron curvature for BL Lacs $k_{\rm BL Lacs}=2.96\pm0.10$ is close to 10/3, which can be explained by statistical particle acceleration for the case of fluctuation of the fractional acceleration gain. The slope of the relation  for radio galaxies $k_{\rm radio, galaxies}=2.23\pm1.38$ is close to 2, which can also be explained by stochastic particle acceleration.      

We find that the slope of the relationship between the synchrotron peak frequency and curvature of different AGNs is different. \cite{Tramacere2011} suggested that the acceleration and cooling processes may explain the difference of the slope. During acceleration, the energy gain has significant dispersion, leading to a momentum diffusion term, i.e., reducing the curvature (namely increasing of $1/b_{\rm sy}$) causes the peak frequency to shift towards higher peak frequencies. When cooling dominates the acceleration process, the slope of this correlation may change  \citep{Tramacere2011, Kalita2019}. \cite{Tramacere2011} thought that magnetic field plays an important role in the evolution of spectral parameters. They proposed that when the magnetic field is weak, the evolution of the particles around the peak is mainly accelerated, while the strong magnetic field is driven by cooling. The FSRQs have a slightly larger magnetic field $B$ (1-10G) than in BL Lacs (0.1-1G) \citep{Ghisellini2010}. The peak of the magnetic field of radio galaxies (M87) is $B=1.61~G$ \citep{Fraija2016}. The average magnetic field
derived for $\gamma$NLS1s is $B=0.91$G \citep{Paliya2019}. The FSRQs and radio galaxies have a slightly larger magnetic field than BL Lacs and $\gamma$NLS1s. These results may indicate that for FSRQs and radio galaxies, cooling dominates the acceleration process, while for BL Lacs and $\gamma$NLS1s, acceleration dominates the cooling process.     

\begin{figure}
	\includegraphics[width=8.5cm,height=8.5cm]{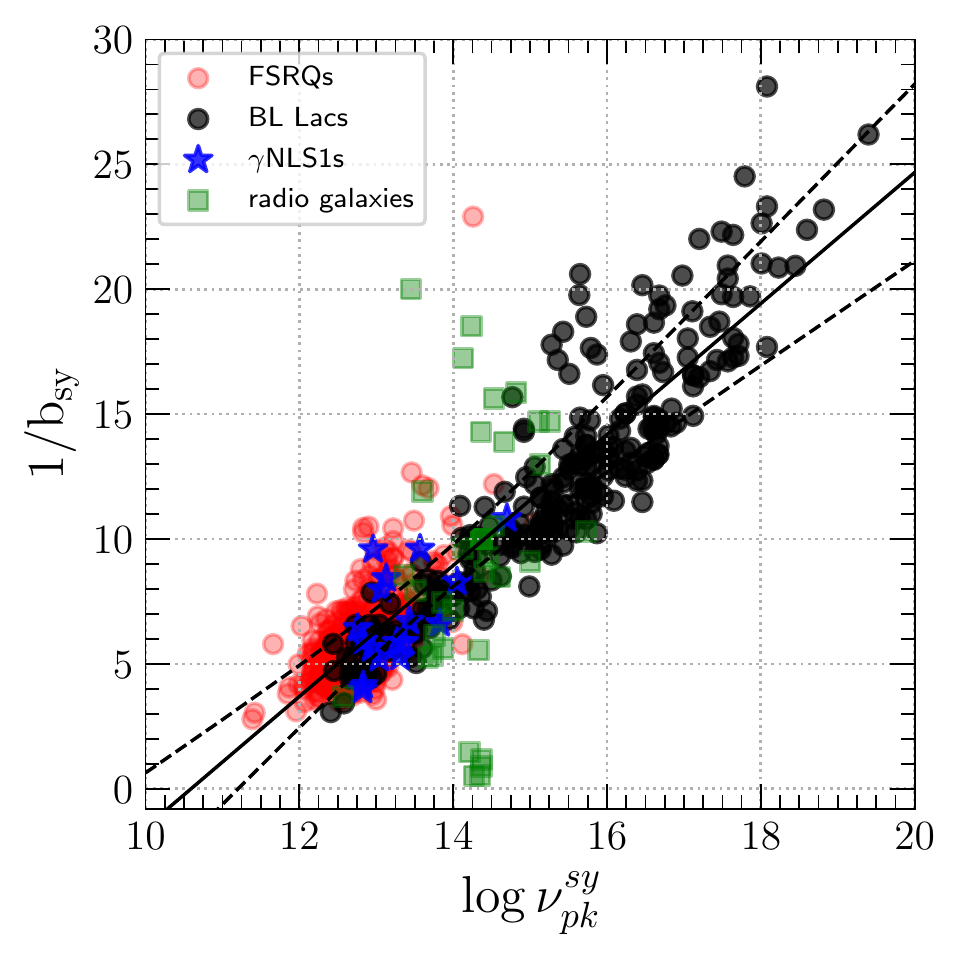}
	\centering
	\caption{Relation between $\log\nu_{\rm pk}^{\rm sy}$ and $1/b_{\rm sy}$ for the whole sample. The solid lines and the dashed lines is the same as figure 3.}
	\label{figure4}
\end{figure}

\subsection{Black hole spins}
We have searched the literature for the most advanced model that can explain the emission of relativistic jets from thin accretion disks, which is applicable to jetted AGNs. The most promising model is the the General relativistic magnetohydrodynamic (GRMHD) simulations of thin the magnetically arrested disks by \cite{Avara2016}.  
\cite{Avara2016} derived an empirical expression of the jet production efficiency based on the thin MAD model,

\begin{figure}
	\includegraphics[width=8.5cm,height=8.5cm]{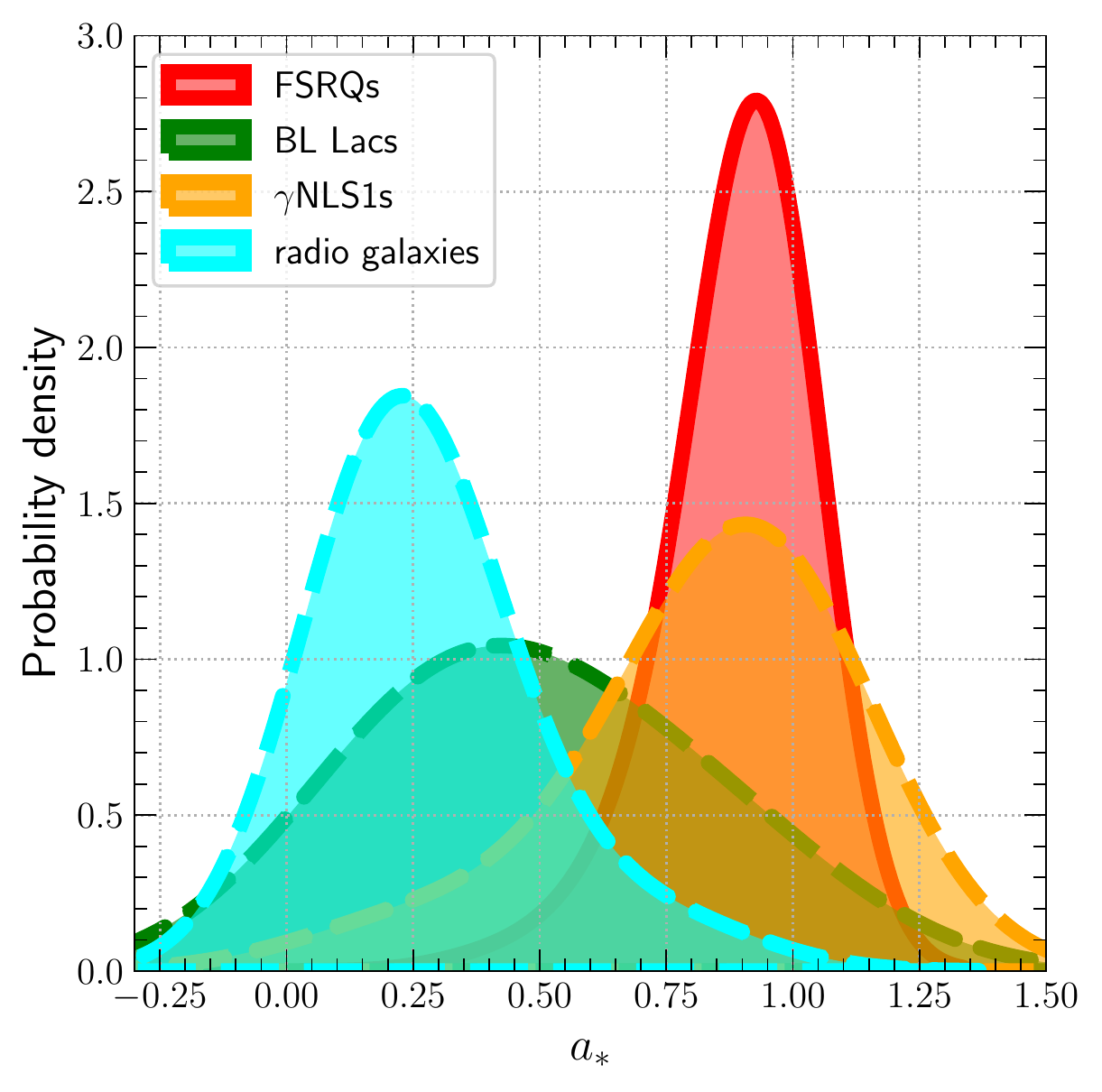}
	\centering
	\caption{Distribution of spins. The red solid histogram is FSRQs. The green dashed histogram is BL Lacs. The orange dashed line is $\gamma$NLS1s. The cyan dashed histogram is radio galaxies.}
	\label{figure5}
\end{figure}    

\begin{equation}
	\eta_{\rm model} \approx 4\omega_{\rm H}^{2}\left(1+\frac{0.3\omega_{\rm H}}{1+2h^{4}}\right)^{2}h^{2},
\end{equation} 
among $\omega_{\rm H}\equiv a_{*}/r_{\rm H}$ is the rotation frequency of the black hole, $r_{\rm H}=1+\sqrt{1-a_{*}^{2}}$ is the event horizon radius, and $h \approx H/R$ is disk thickness \citep{McKinney2012, McKinney2013, Avara2016}. \cite{Soares2020} solved equation (12) to obtain the spin of a black hole by assuming $\eta = \eta_{\rm model}$, $\eta=P_{\rm jet}/(\dot{Mc^{2}})$, $\dot{M}=0.1\dot{M_{\rm Edd}}=L_{\rm Edd}/c^{2}$, and $h=0.13$. The observed jet efficiency should satisfy the following requirements,

\begin{equation}
	\eta < max(\eta_{\rm model}) = \eta_{\rm model}(a_{*} = 0.998) = 0.098, 
\end{equation}
where the maximum allowed spin is $max(a)=0.998$ \citep{Thorne1974}. Following the work of \cite{Soares2020}, we also use equation (12) to get the black hole spins for our sample with $\eta < max(\eta_{\rm model})$. 

Figure 6 shows the distributions of black hole spin for FSRQs, $\gamma$NLS1s, BL Lacs, and radio galaxies. The average values of black hole spin for FSRQs are $a_{*}=0.93\pm0.11$, for BL Lacs are $a_{*}=0.44\pm0.26$, for $\gamma$NLS1s are $a_{*}=0.91\pm0.22$, and for radio galaxies are $a_{*}=0.23\pm0.17$. \cite{Soares2020} got that the average values of black hole spin for 154 FSRQs are $a_{*}=0.84_{-0.25}^{+0.11}$. Our results are slightly different from theirs. There are two possible reasons. One is the number of objects, our number of objects is larger than theirs. Another reason is that the calculation method of jet power is different. They mainly use gamma-ray luminosity to get jet power, and we use the 1.4 GHz radio luminosity. We find that the FSRQs and $\gamma$NLS1s have high black hole spin than BL Lacs and radio galaxies. The anti-correlation between jet kinetic power and the synchrotron peak frequency reflects the existence of the blazar sequence. Thus, the relation between the spin of black hole and the synchrotron peak frequency also can be used to test the Fermi blazar sequence.         

The relation between the spin of black and the synchrotron peak frequency is shown in Figure 7. The Pearson analysis shows a significant anti-correlation between black hole spin and the synchrotron peak frequency for the whole sample ($r=-0.71$, $P=1.35\times10^{-118}$). The Spearman correlation coefficient and significance level for this relation are $r=-0.68$ and $P=3.21\times10^{-102}$. The Kendall tau correlation coefficient and significance level for this relation are $r=-0.45$ and $P=1.36\times10^{-77}$. The correlation between black hole spin and the synchrotron peak frequency is significant even after the effects of jet kinetic power
are removed ($r_{\rm XY, Pjet}=-0.35$, $P=1.09\times10^{-22}$). The fitted equation is as follows,

\begin{equation}
	a_{*}=(-0.129\pm0.005)\log\nu_{\rm pk}^{\rm sy}+(2.493\pm0.065).
\end{equation} 
To sum up, all the results and discussions support the Fermi blazar sequence. The $\gamma$NLS1s and radio galaxies belong to the Fermi blazar sequence. \cite{Paliya2013} suggested that the physical properties of the two gamma ray emitting NLSy1 galaxies (PKS1502+036 and PKS2004-447) are similar to blazar, between FSRQs and BLLac objects, so these sources may belong to the traditional blazar sequence. 

\begin{figure}
	\includegraphics[width=8.5cm,height=8.5cm]{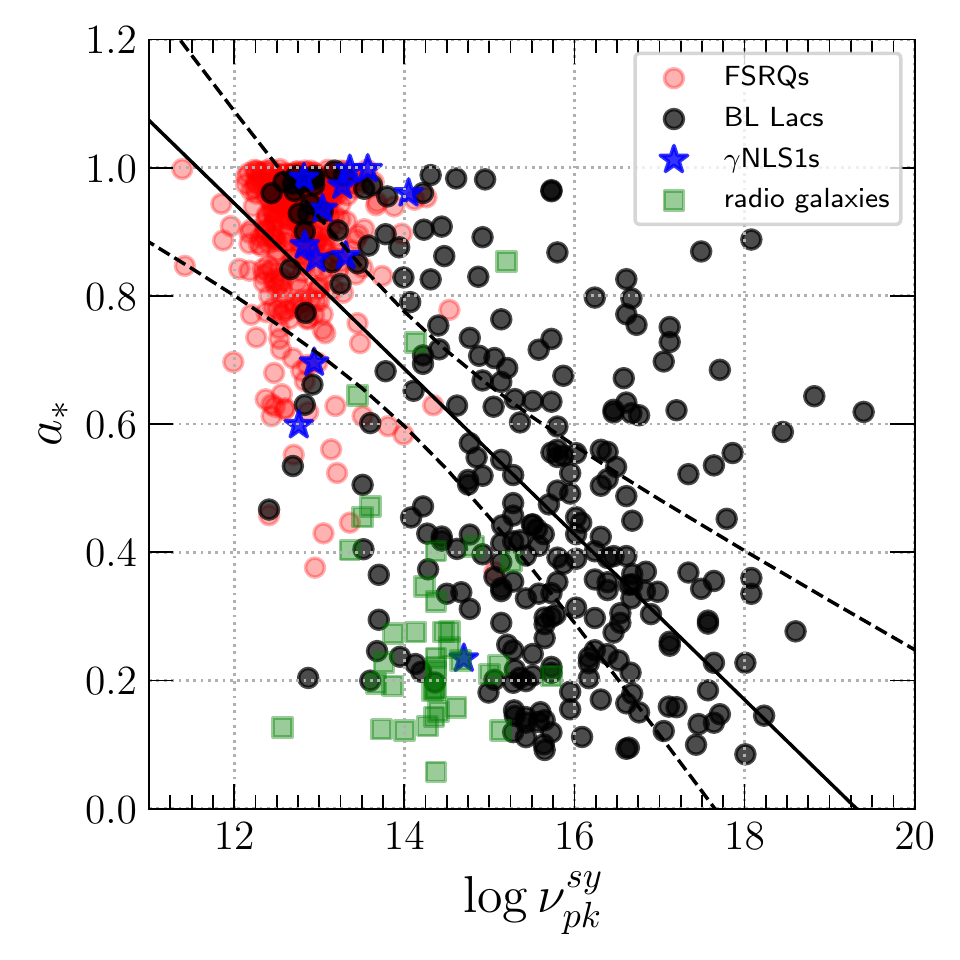}
	\centering
	\caption{Relation between black hole spin and the synchrotron peak frequency for the whole sample. The solid lines and the dashed lines is the same as figure 3.}
	\label{figure6}
\end{figure}

\section{Conclusions}
We use the largest Fermi samples, including Fermi blazars, $\gamma$NLS1s, and radio galaxies, to study the relationship between them. Our main results are as follows,

(1) In the plane of the synchrotron peak frequency luminosity, jet kinetic power, $\gamma$-ray luminosity, and the synchrotron peak frequency, the Fermi blazar sequence is ``L'' shaped.  

(2) There is a significant inverse correlation between Compton dominance, black hole spin and the synchrotron peak frequency for the whole sample. These results support the Fermi blazar sequence, and $\gamma$NLS1s and radio galaxies belong to the Fermi blazar sequence. 

(3) There is a significant correlation between synchrotron
peak frequency and synchrotron curvature for the whole
sample, Fermi blazars, FSRQs, and $\gamma$NLS1s, respectively. The slopes of such a correlation are close to 2.5, which implies that these AGNs can be explained by statistical acceleration for the case of energy-dependent acceleration probability. However, the slope of such a correlation for BL Lacs is close to 10/3, which implies that BL Lacs can be explained by statistical particle acceleration in the case of fluctuation of the fractional acceleration gain. The slope of this correlation for radio galaxies is close to 2, which implies that radio galaxies can be explained by stochastic particle acceleration.           

(4) The average value of black hole spin for FSRQs is $a_{*}=0.93\pm0.11$, for BL Lacs is $a_{*}=0.44\pm0.26$, for $\gamma$NLS1s is $a_{*}=0.91\pm0.22$, for radio galaxies is $a_{*}=0.23\pm0.17$. 

\acknowledgments We are very grateful to the referee and Editor for the very helpful comments and suggestions that improved the presentation of the manuscript substantially. Yongyun Chen is gratefulfor financial support from the National Natural Science Foundation of China (No. 12203028). This work was support from the research project of Qujing Normal University (2105098001/094). This work is supported by the youth project of Yunnan Provincial Science and Technology Department (202101AU070146, 2103010006). Yongyun Chen is grateful for funding for the training Program for talents in Xingdian, Yunnan Province. 
QSGU is supported by the National Natural Science Foundation of China (No. 11733002, 12121003, 12192220, and 12192222).
We also acknowledge the science research grants from the China Manned Space Project with NO. CMS-CSST-2021-A05. This work is supported by the National Natural Science Foundation of China (11733001 and U2031201).

%Need likeapj.bst in directory
\bibliographystyle{aasjournal}
%Need example.bib in directory 
\bibliography{example}    

\label{lastpage}

\end{document}